\documentclass[a4paper,12pt]{article}

\usepackage{amsfonts,amsmath}
\usepackage{amssymb}
\usepackage{amscd}

\usepackage[mathscr]{eucal}

\usepackage{amsthm}
\newtheorem{theorem}{Theorem}[section]
\newtheorem{definition}[theorem]{Definition}

\def\endproof{\ \hfill\hbox{\vbox{\hrule\hbox{\vrule
height5pt\kern5pt\vrule height5pt}\hrule}}\par\medskip\rm}
\evensidemargin \oddsidemargin 
\newcommand{\be}{\begin{equation}}
\newcommand{\ee}{\end{equation}}

\bigskip

\title{{
\bf 
Periodic coherent states decomposition and quantum dynamics on the flat torus  
}
}

\author{  Lorenzo Zanelli\footnote{ lzanelli@math.unipd.it} \\ Department of Mathematics ``Tullio Levi-Civita'' \\ University of Padova  }

\date{}

\begin{document}

\maketitle

\begin{abstract}
\noindent
We provide a result on the coherent states decomposition for functions in $L^2 (\Bbb T^n)$  where $\Bbb T^n :=  (\Bbb R / 2\pi \Bbb Z)^n$ is the flat torus. Moreover, we  study  such a decomposition with respect to the quantum dynamics related to semiclassical elliptic Pseudodifferential operators, and we prove a related invariance result.
\end{abstract}

{\bf keywords}: Coherent states, toroidal Pdo, quantum dynamics.

%
%
%
%
%
%
%
%
%





\section{Introduction}
Let us introduce the usual class of semiclassical coherent states on $\Bbb R^n$ 
\begin{equation}
\label{cohe-1}
\phi_{(x,\xi)} (y) := \alpha_h \, e^{ \frac{i}{h} (x-y) \cdot \xi } e^{- \frac{|x-y|^2}{2h} }, \quad (x,\xi) \in \Bbb R^{2n}, \quad y \in \Bbb R^n, \quad 0 < h \le 1
\end{equation}
with the $L^2 (\Bbb R^n)$ - normalization constant $\alpha_h := 2^{- \frac{n}{2}}  (\pi h)^{-3n/4}$, and where $h$ is a `semiclassical parameter'. For any $\psi \in \mathcal{S}^\prime (\Bbb R^n)$ the coherent state decomposition reads, in the distributional sense, as 
\begin{equation}
\label{deco-1}
\psi (x_0)  =   \int_{\Bbb R^{2n}}  \phi_{(x,\xi)}^\star (x_0)  \Big( \int_{\Bbb R^{n}}      \phi_{(x,\xi)} (y)     \psi (y) dy  \Big)   dxd\xi
\end{equation}
as shown  for example in Prop. 3.1.6 in \cite{Mart}.\\
 We now observe that for the flat torus $\Bbb T^n :=  (\Bbb R / 2\pi \Bbb Z)^n$  the well known inclusion   $L^2 (\Bbb T^n) \subset \mathcal{S}^\prime (\Bbb R^n)$  implies that distributional equality (\ref{deco-1}) make sense also for functions in  $L^2 (\Bbb T^n)$.\\
The first aim of our paper is to prove the decomposition of any $\varphi \in L^2 (\Bbb T^n)$ with respect to the family of periodic coherent states  $\Phi$ given by the periodization of (\ref{cohe-1}). In view of this target, we recall that the periodization operator $\Pi (\phi) (y) := \sum_{k \in \Bbb Z^n} \phi ( y -  2\pi k )$ maps $\mathcal{S} (\Bbb R^n)$ into $C^\infty (\Bbb T^n)$, as shown for example Thm. 6.2 in \cite{R-T}. 
Thus, we can define  for all  $0 < h \le 1$ 
\begin{equation}
\label{cohe-2}
\Phi_{(x,\xi)} (y) :=  \sum_{k \in \Bbb Z^n} \phi_{(x,\xi)} (y - 2\pi k) \qquad (x,\xi) \in \Bbb T^{n} \times h \, \Bbb Z^n, \qquad y \in \Bbb T^n.
\end{equation}
\noindent
Notice that the family of coherent states  in (\ref{cohe-2}) is well posed also for $\xi \in \Bbb R^n$ and the related phase space is $\Bbb T^{n} \times  \Bbb R^n$. However, our target is to show that the decomposition of periodic functions can be done with respect to the minimal set of coherent states in  (\ref{cohe-2}) for $\xi \in  h \, \Bbb Z^n  \subset \Bbb R^n$. Furthermore, we notice that the phase space $\Bbb T^{n} \times h \, \Bbb Z^n$ is necessary in order to deal with a well defined setting of toroidal Weyl operators acting on $L^2 (\Bbb T^n)$ and more in general with semiclassical toroidal Pseudodifferential operators (see Sect. \ref{SEC2}).

The first result of the paper is the following
\begin{theorem}
\label{TH1}
Let $\varphi_h \in C^\infty (\Bbb T^n)$, $\|  \varphi_h \|_{L^2} = 1$ with $0 < h \le 1$, 
$h^{-1} \in \Bbb N$  and let $\Phi_{(x,\xi)}$ be as in (\ref{cohe-2}). Then, 
\begin{equation}
\label{deco-02}
\varphi_h     = \sum_{\xi \in h \, \Bbb Z^n} \int_{\Bbb T^n}   \langle \Phi_{(x,\xi)} , \varphi_h   \rangle_{L^2 } \, \Phi_{(x,\xi)}^\star   \, dx +  \mathcal{O}_{L^2} (h^\infty).
\end{equation}
Moreover, there exists $f(h) > 0$ depending on $\varphi_h$ such that  
\begin{equation}
\label{psi-dec00}
\varphi_h  = \sum_{\xi \in h \, \Bbb Z^n, \ |\xi| \le f(h) } \ \int_{\Bbb T^n}   \langle \Phi_{(x,\xi)} , \varphi_h   \rangle_{L^2 } \, \Phi_{(x,\xi)}^\star  \, dx   +   \mathcal{O}_{L^2} (h^\infty).
\end{equation}
\end{theorem}   
\noindent

The following inclusion involving  the  set of frequencies $\xi \in h \, \Bbb Z^n \subset \Bbb R^n$ allows to consider decomposition (\ref{deco-02})  minimal  with respect to    (\ref{deco-1}). The above result shows also that  the sum over the frequencies can be  taken in the bounded  region $|\xi| \le f(h)$, i.e. we can consider a finite sum by taking into account an  $\mathcal{O}(h^\infty)$ remainder in $L^2 (\Bbb T^n)$. \\An analogous result of (\ref{deco-02})  in the two dimensional setting is shown in prop. 60 of \cite{comb-rob} by the use of a different periodization operator. Same construction of coherent states as in \cite{comb-rob} for $\Bbb T^2$ is used in \cite{Bo-Bi}, \cite{Fa-N-B} for the study of quantum cat maps and equipartition of the eigenfunctions of quantized ergodic maps. In the paper \cite{FGN}, covariant integral quantization using coherent states for semi-direct product groups is implemented for the motion of a particle on the circle and in particular the resolution of the identity formula is proved. Another class of coherent states on the torus are defined also in \cite{Fre}, with a related resolution of the identity, in the understanding of the Quantum Hall effect.  We also recall \cite{D-B} where coherent states and Bargmann Transform are studied on $L^2 (\Bbb S^n)$.
The literature on coherent states are quite rich, and thus we address the reader  to \cite{A-B-G}.\\
We now devote our attention to the periodic coherent states decomposition for eigenfunctions of elliptic semiclassical toroidal Pseudodifferential operators (see Section  \ref{SEC2}). We will see that  the formula (\ref{deco-02}) can be reduced in view of a phase-space localization of eigenfunctions.\\
\indent This is the content of the second main result of the paper.

\begin{theorem}
\label{TH2}
Let $\mathrm{Op}_h (b)$ be an elliptic semiclassical $\Psi$do as in (\ref{sem-pdo}) and $h^{-1} \in \Bbb N$. Let $E \in \Bbb R$, and let $\psi_h \in C^\infty (\Bbb T^n)$,   $\|  \psi_h \|_{L^2} = 1$ and which is eigenfunction of the eigenvalue problem on $\Bbb T^n$  
\begin{equation*}
\label{Eige-01}
\mathrm{Op}_h (b) \psi_h = E_h \psi_h 
\end{equation*}
where $E_h \le E$ for any $0 < h \le 1$. Then, there exists  $g (h,E) \in \Bbb R_+$ such that 
\begin{equation}
\label{deco-03}
\psi_h  = \sum_{\xi \in h \, \Bbb Z^n, \ |\xi| \le g(h,E) } \ \int_{\Bbb T^n}   \langle \Phi_{(x,\xi)} , \psi_h   \rangle_{L^2 } \, \Phi_{(x,\xi)}^\star  \, dx   +   \mathcal{O}_{L^2 } (h^\infty).
\end{equation}
\end{theorem}  
We notice that for the operators $- h^2 \Delta_x + V(x)$ all the eigenfunctions with eigenvalues $E_h \le E$ fulfill  $\|\Delta_x \psi_h \|_{L^2} \le c \, h^{-2}$. 
In particular, we have the asymptotics $g (h,E) \to + \infty$ as $h \to 0^+$.  We also underline that the function $g(E,h)$ and the estimate on remainder $\mathcal{O}_{L^2 }(h^\infty)$  do not depend on the particular choice of $\psi_h$. This implies that all these eigenfunctions take the form (\ref{deco-03}) and therefore also any finite linear combination of eigenfunctions of kind $\sum_{1 \le \alpha \le N} c_\alpha \, \psi_{h,\alpha}$ where $|c_\alpha| \le 1$. We remind that  Weyl Law on the number $\mathcal{N}(h)$ of eigenvalues $E_{h,\alpha} \le E$ (with their multiplicity)  for semiclas\-sical elliptic operators (see for example \cite{G-S}) reads $\mathcal{N}(h) \simeq  (2\pi h)^{-n}( {\rm vol}(U(E))  + \mathcal{O}(1) )$.\\  
 The proof of the above result is mainly based  on a uniform estimate for our toroidal version of the  Fourier-Bros-Iagolnitzer (FBI) transform
\begin{equation*}
( \mathcal{T} \psi_h ) (x,\xi)  :=  \langle \Phi_{(x,\xi)} , \psi_h   \rangle_{L^2} 
\end{equation*}
on the unbounded region given by  all  $x \in \Bbb T^n$  and  $\xi \in h \Bbb Z^n$ such that $|\xi| > g(h,E)$.  The FBI transform on any compact manifold has already been defined and studied  in the literature, see for example \cite{Wu-Z}. \\
 We remind that, in the euclidean setting of $\Bbb R^{2n}$, the function defined as $T_h  (\psi_h) (x,\xi) :=  \langle \phi_{(x,\xi)} , \psi_h   \rangle_{L^2 (\Bbb R^n)}$ is the usual version of the  
FBI transform, which is well posed for any  $\psi_h \in \mathcal{S}^\prime (\Bbb R^n)$. This is used to study the phase space localization by the Microsupport of $\psi_h$ (see for example \cite{Mart}), namely $MS(\psi_h)$ the complement of the set of points $(x_0,\xi_0)$  such that  $T_h  (\psi_h) (x,\xi) \simeq \mathcal{O}( e^{- \delta / h})$ uniformly in a neighborhood of $(x_0,\xi_0)$. 
In the case of the weaker estimate $T_h  (\psi_h) (x,\xi) \simeq \mathcal{O}(h^\infty)$ one can define the semiclassical Wave Front Set $WF(\psi_h)$.  
 Is well known (see \cite{Mart}) that the Microsupport (or the semiclassical Wave Front Set) of eigenfunctions for elliptic operators is localized in the  sublevel sets  $U (E) := \{ (x,\xi) \in \Bbb R^{2n} \ | \  b(x,\xi) \le E \}$, i.e. $MS(\psi_h) \subseteq U(E)$. 
 The well posedness of $WF (\psi_h)$ and  $MS(\psi_h)$ in the periodic setting can be seen starting from the euclidean setting and thanks to distributional inclusion $L^2 (\Bbb T^n) \subset \mathcal{S}^\prime (\Bbb R^n)$, (see for example section 3.1 of \cite{C-Z}). The semiclassical study in the phase space for eigenfunctions in the periodic setting has also been studied in \cite{Z2} with respect to weak KAM theory.\\ 
 In our Theorem \ref{TH2} we are interested to show another kind of semiclassical localization, namely to localize the bounded region $\Omega(E,h) := \{ (x,\xi) \in \Bbb T^n \times \Bbb R^n \ | \  x \in \Bbb T^n, |\xi| \le g(E,h) \}$ which will be bigger than $MS(\psi_h)$,  $h$ - dependent and such that the coherent state decomposition of $\psi_h$ can be done up to a remainder $\mathcal{O}_{L^2 }(h^\infty)$.

We now focus our attention to the decompositon (\ref{deco-03}) under the time evolution.
\begin{theorem}
\label{TH3}
Let $\varphi_h \in C^\infty (\Bbb T^n)$, $L^2$ - normalized such that 
\begin{equation*}
\label{LL-comb}
\varphi_h  = \sum_{1 \le j \le J (h)} c_j \, \psi_{h,j}
\end{equation*}
 where $\psi_{h,j}$ are given in Thm. \ref{TH2} and $J(h) \le J_0 h^{-Q}$ for some $J_0, Q > 0$. Let $\mathrm{Op}_h (b)$ be an elliptic semiclassical $\Psi$do as in (\ref{sem-pdo}) and $U_h (t) := \exp\{(- i \mathrm{Op}_h (b) t  )  / h \}$. Then, there exists $\ell (h) > 0$ s.t. for any $t \in \Bbb R$ 
\begin{equation}
\label{psi-dec22}
U_h (t) \varphi_h  = \sum_{\xi \in h \, \Bbb Z^n, \ |\xi| \le \ell(h) } \ \int_{\Bbb T^n}   \langle \Phi_{(x,\xi)} ,  U_h (t) \varphi_h   \rangle_{L^2 } \, \Phi_{(x,\xi)}^\star  \, dx   +   \mathcal{O}_{L^2 }(h^\infty).
\end{equation}
\end{theorem} 
\noindent 
The equality (\ref{psi-dec22}) shows that time evolution under the $L^2$ - unitary map $U_h (t)$ does not change such a  decomposition, since $\ell (h)$ does not depend on time. 
The function $\ell (h)$ is not necessarily the same as the function $f(h)$ contained in Theorem \ref{TH1} but we have that $\ell (h) \ge f(h)$.
In other words, this quantum dynamics preserves the coherent state decomposition (\ref{psi-dec00}). The same result holds for any eigenfunctions in Thm. \ref{TH2} since in this case  $U_h (t) \psi_h = \exp\{(- i E_h  t) / h \}  \psi_h$.
Notice that here we can assume that $Q > n$, namely the linear combination (\ref{LL-comb}) can be done with more eigenfunctions  than the ones that have eigenvalues $E_{h} \le E$ with fixed energy $E > 0$. 
 Notice also that  we have  $ \langle \Phi_{(x,\xi)} ,  U_h (t) \varphi_h   \rangle_{L^2 } =  \langle U_h (-t) \Phi_{(x,\xi)} ,   \varphi_h   \rangle_{L^2 }$ for any $t \in \Bbb R$ and that the time evolution of the periodization of coherent states has been used in \cite{Z1} in the context of optimal transport theory.


\section{Semiclassical toroidal Pseudodifferential operators}

\label{SEC2}
Let us define  the flat torus $\Bbb T^n := (\Bbb R / 2\pi \Bbb Z)^n$  and introduce the  class of   symbols $b \in S^m_{\rho, \delta} (\mathbb{T}^n \times \mathbb{R}^n)$, $m \in \mathbb{R}$, $0 \le \delta$, $\rho \le1$, given by functions  
in $C^\infty (\mathbb{T}^n \times \mathbb{R}^n;\Bbb R)$ which are  $2\pi$-periodic in each variable $x_j$, $1\leq j\leq n$ and for which
for all $\alpha, \beta \in \mathbb{Z}_+^n$ there exists $C_{\alpha \beta} >0$ such that $\forall$ $(x,\xi) \in \mathbb{T}^n \times \mathbb{R}^n$
\begin{equation*}
\label{symb00}
|  \partial_x^\beta \partial_\xi^\alpha  b (x,\xi)  |   \le  C_{\alpha \beta m} \langle \xi \rangle^{m- \rho |\alpha| + \delta |\beta|}
\end{equation*}
where $\langle\xi\rangle:=(1+|\xi|^2)^{1/2}$. In particular,  the set $S^m_{1,0} (\mathbb{T}^n \times \mathbb{R}^n)$ is denoted by $S^m (\mathbb{T}^n \times \mathbb{R}^n)$.\\ 
We introduce the semiclassical toroidal Pseudodifferential Operators by 
\begin{definition}
Let $\psi \in C^\infty (\mathbb{T}^n;\Bbb C)$ and $0 < h \le 1$, 
\begin{equation*}
\label{sem-pdo}  
\mathrm{Op}_h(b) \psi(x):=(2\pi)^{-n}\sum_{\kappa \in\mathbb{Z}^n}\int_{\mathbb{T}^n}e^{i\langle x-y,\kappa \rangle}b(x,h \kappa)\psi(y)dy.
\end{equation*}
\end{definition}
This is the semiclassical version (see \cite{P-Z}, \cite{TP-Z}) of the quantization by Pseudodifferential Operators on the torus developed in \cite{R-T} and \cite{R-T-book}. See also \cite{B-D} for the notion of vector valued  Pseudodifferential Operators on the torus. 
\noindent

We now notice that we have  have  a map $\mathrm{Op}_h(b) : C^\infty (\mathbb{T}^n) \longrightarrow \mathcal{D}^\prime  (\mathbb{T}^n)$.  Indeed, remind that  $u \in \mathcal{D}^\prime  (\mathbb{T}^n)$ are the linear maps $u: C^\infty (\mathbb{T}^n) \longrightarrow \Bbb C$ such that $\exists$ $C>0$ and $k \in \Bbb N$, for which $|u(\phi)| \le C \sum_{|\alpha|\le k} \| \partial_x^\alpha \phi  \|_\infty$ $\forall \phi \in C^\infty (\Bbb T^n)$.\\ 
Given a symbol $b\in S^m(\mathbb{T}^n\times\mathbb{R}^n)$, the toroidal Weyl quantization reads (see Ref. \cite{P-Z}, \cite{TP-Z})
\begin{equation*}
\label{weyl}
\mathrm{Op}^w_h(b) \psi(x) := (2\pi)^{-n}\sum_{\kappa \in\mathbb{Z}^n}\int_{\mathbb{T}^n}e^{i\langle x-y,\kappa \rangle}b \Big(y, \frac{h}{2} \kappa \Big)\psi(2y-x)dy.
\end{equation*}
In particular, it  holds  
\begin{equation*}
\label{eq-O}
\mathrm{Op}^w_{h} (b)  \psi (x) = (\mathrm{Op}_h(\sigma) \circ T_x \, \psi )(x)
\end{equation*}
where  $T_x : C^\infty (\mathbb{T}^n) \rightarrow C^\infty (\mathbb{T}^n)$ defined as $(T_x \psi) (y) := \psi (2y-x)$  is linear, invertible and $L^2$-norm preserving, and $\sigma$ is a suitable toroidal symbol related to $b$, i.e.  $\sigma \sim \sum_{\alpha\geq 0}\frac{1}{\alpha!}\triangle_\xi^\alpha D_y^{(\alpha)} b(y,h \xi / 2)\bigl|_{y=x}$, where $\triangle_{\xi_j} f (\xi + e_j) - f(\xi)$ is the difference operator (see Th. 4.2 in Ref. \cite{R-T}).\\ 
The typical example is given by 
\begin{eqnarray*}
\mathrm{Op}_h (H)  &=& \Big(- \frac{1}{2} h^2 \Delta_x + V (x) \Big)\psi(x) \\
&=& (2\pi)^{-n}\sum_{\kappa \in\mathbb{Z}^n}\int_{\mathbb{T}^n}e^{i\langle x-y,\kappa \rangle}  \Big(\frac{1}{2}|h \kappa|^2  + V(x) \Big)  \psi(y)dy
\end{eqnarray*}
namely the related symbol is the mechanical type Hamiltonian $H (x,\xi) = \frac{1}{2} |\xi|^2 + V(x)$. Also in the case of the Weyl operators we have 
\begin{equation*}
- \frac{1}{2} h^2 \Delta_x + V (x) =  \mathrm{Op}^w_{h} (H)
\end{equation*}
for the same symbol (see for example \cite{TP-Z}).\\
In our paper we are interested in uniform elliptic operators, namely such that the symbol $b \in S^m (\mathbb{T}^n \times \mathbb{R}^n)$ fulfills for some constants $C,c > 0$ the lower bound
\begin{equation*}
|b(x,\xi)| \ge C \, \langle \xi \rangle^m
\end{equation*}
for any $x \in \Bbb T^n$  and $|\xi| \ge c$. This property guarantees bounded sublevels sets for $b$ and discrete spectrum for the operator $\mathrm{Op}_h(b)$ for any fixed $0 < h \le 1$.
As we see in Theorem \ref{TH2}, this assumption permits  also to prove the semiclassical localization of all the eigenfunctions within these sublevels sets, and this localization can be studied by  our semiclassical coherent states (\ref{cohe-2}).

\section{Main Results}

\noindent
{\bf Proof of Theorem 1.1}
We remind that  $\Phi_{(x,\xi)} (y) := \Pi (\phi_{(x,\xi)}) (y)$  and $\Pi (\phi) (y) := \sum_{k \in \Bbb Z^n} \phi ( y -  2\pi k )$. Thus,
\begin{eqnarray*}  
\Phi_{(x + 2\pi \beta,\xi)} (y)  &=&  \sum_{k \in \Bbb Z^n} \phi_{(x + 2\pi \beta,\xi)}  (y - 2\pi k)  =   \sum_{k \in \Bbb Z^n} \phi_{(x ,\xi)}  (y - 2\pi k - 2\pi \beta)
\nonumber
\\
&=&  \Phi_{(x,\xi)} (y).  
\end{eqnarray*} 
We mainly adapt, in our toroidal setting,  the proof of Prop. 3.1.6 shown in \cite{Mart} written for the euclidean setting. Thus, 
we define the operator $\mathcal{T}^\star$ on functions $\Psi  \in L^2 ( \Bbb T^n \times h \Bbb Z^n )$ as 
\begin{equation*} 
( \mathcal{T}^\star \Psi ) (y)  :=  \sum_{\xi \in h \, \Bbb Z^n} \ \int_{\Bbb T^n}  \Psi (x, \xi) \Phi_{(x,\xi)}^\star (y)  \, dx. 
\end{equation*} 
It can be easily seen that $\mathcal{T}^\star$ equals the adjoint of the operator  $( \mathcal{T} \psi ) (x,\xi)  :=  \langle \Phi_{(x,\xi)} , \psi   \rangle_{L^2 (\Bbb T^n)}$, i.e.
\begin{equation*} 
\langle  \mathcal{T}^\star \Psi ,  \psi \rangle_{L^2 (\Bbb T^n)} =  \langle  \Psi , \mathcal{T}  \psi \rangle_{L^2 (\Bbb T^n \times h \Bbb Z^n)}. 
\end{equation*}  
Thus, $\forall \psi_1, \psi_2 \in C^\infty (\Bbb T^n) \subset L^2 (\Bbb T^n)$ we have
\begin{equation*} 
\langle  \mathcal{T}^\star \circ \mathcal{T} \psi_1 ,  \psi_2 \rangle_{L^2 (\Bbb T^n)} =  \langle  \mathcal{T} \psi_1 , \mathcal{T}  \psi_2 \rangle_{L^2 (\Bbb T^n \times h \Bbb Z^n)}. 
\end{equation*} 
It remains to prove that
\begin{equation} 
\label{orto-per}
\langle  \mathcal{T} \psi_1 , \mathcal{T}  \psi_2 \rangle_{L^2 (\Bbb T^n \times h \Bbb Z^n)} =  \langle  \psi_1 ,   \psi_2 \rangle_{L^2 (\Bbb T^n)} + \mathcal{O}(h^\infty)
\end{equation}
which implies 
\begin{equation} 
\label{TT}
\mathcal{T}^\star \circ \mathcal{T} = {\rm Id} \ {\rm mod} \ \mathcal{O}(h^\infty)
\end{equation} 
on $L^2 (\Bbb T^n)$, and equality (\ref{TT}) is exactly the statement  (\ref{deco-02}).\\
In order to prove (\ref{orto-per}), we recall  that the  periodization operator $\Pi$ can be rewritten in the form (see Thm. 6.2 in \cite{R-T}):
 \begin{equation} 
 \label{per-op1}
\Pi (\phi)  = \mathcal{F}_{\Bbb T^n}^{-1}  \Big(   \mathcal{F}_{\Bbb R^n}  \phi \Big|_{\Bbb Z^n} \Big). 
\end{equation}  
where $ \mathcal{F}_{\Bbb T^n}^{-1}$ stands for the inverse toroidal Fourier Transform, and $\mathcal{F}_{\Bbb R^n}$ is the usual euclidean version.
In view of (\ref{per-op1}) it follows
 \begin{eqnarray*} 
( \mathcal{T} \psi ) (x,\xi)  &:=&  \langle \Phi_{(x,\xi)} , \psi   \rangle_{L^2 (\Bbb T^n)} =  \langle  \mathcal{F}_{\Bbb R^n}  \phi_{x,\xi} |_{\Bbb Z^n} , \mathcal{F}_{\Bbb T^n} \psi   \rangle_{L^2 (\Bbb Z^n)} 
\\
&=&   \sum_{ k \in \Bbb Z^n}   \widehat{\phi}_{x,\xi}  (k)^\star  \widehat{\psi} (k) ,
\end{eqnarray*}  
where  $ \widehat{\phi}_{x,\xi}  (k)  := \mathcal{F}_{\Bbb R^n}  \phi_{x,\xi} (k)$ and  $\widehat{\psi} (k) :=  \mathcal{F}_{\Bbb T^n} \psi (k)$.
Thus,
 \begin{eqnarray*} 
&& \langle   \mathcal{T} \psi_1 ,    \mathcal{T}  \psi_2 \rangle_{L^2 (\Bbb T^n \times h \Bbb Z^n)}
\\
&& =   \sum_{\xi \in h \, \Bbb Z^n} \int_{\Bbb T^n}   \Big(  \sum_{ k \in \Bbb Z^n}   \widehat{\phi}_{x,\xi}  (k)^\star  \widehat{\psi} (k)  \Big)^\star   \Big(  \sum_{ \mu \in \Bbb Z^n}   \widehat{\phi}_{x,\xi}  (\mu)^\star  \widehat{\psi} (\mu)  \Big) dx .
 \end{eqnarray*} 
We can rewrite this equality, in the distributional sense, as
 \begin{eqnarray*} 
&&  \langle   \mathcal{T} \psi_1 ,    \mathcal{T}  \psi_2 \rangle_{L^2 (\Bbb T^n \times h \Bbb Z^n)}  
\\
&&  =  \sum_{ k,\mu \in \Bbb Z^n}   \widehat{\psi}_1 (k)^\star \widehat{\psi}_2 (\mu)   \sum_{\xi \in h \, \Bbb Z^n} \int_{Q}       \widehat{\phi}_{x,\xi}  (k)     \widehat{\phi}_{x,\xi}  (\mu)^\star  dx 
 \end{eqnarray*} 
where $Q :=  [0,2\pi]^n$ and $\psi_1, \psi_2 \in C^\infty (\Bbb T^n)$. Now let $\xi = h \alpha$ with $\alpha \in \Bbb Z^n$, so  that
 \begin{eqnarray*} 
&&  \langle   \mathcal{T} \psi_1 ,  \mathcal{T}  \psi_2 \rangle_{L^2 (\Bbb T^n \times h \Bbb Z^n)}  
\\
&&  =  \sum_{ k,\mu \in \Bbb Z^n}   \widehat{\psi}_1 (k)^\star \widehat{\psi}_2 (\mu)   \sum_{\alpha \in  \Bbb Z^n} \int_{Q}       \widehat{\phi}_{x,h \alpha}  (k)     \widehat{\phi}_{x,h \alpha}  (\mu)^\star  dx 
 \end{eqnarray*} 
 By using the explicit form of $\widehat{\phi}_{x,h \alpha}$ and the condition $h^{-1} \in \Bbb N$, a direct computation shows that
 \begin{eqnarray} 
&&  \langle   \mathcal{T} \psi_1 ,  \mathcal{T}  \psi_2 \rangle_{L^2 (\Bbb T^n)}  
\nonumber
\\
&&  =  \sum_{ k,\mu \in \Bbb Z^n}   \widehat{\psi}_1 (k)^\star \widehat{\psi}_2 (\mu) \Big[ \Big( \sum_{\alpha \in  \Bbb Z^n} e^{i \alpha (k-\mu)} \Big) + \mathcal{O}(h^\infty)  \Big]
\label{qq-cat}
 \end{eqnarray} 
 where $\mathcal{O}(h^\infty)$ does not depend on the functions $\psi_1,\psi_2$.   
 To conclude, since $\delta (k-\mu) =  \sum_{\alpha \in  \Bbb Z^n} e^{i \alpha (k-\mu)}$, we get
 \begin{eqnarray} 
  \langle   \mathcal{T} \psi_1 ,  \mathcal{T}  \psi_2 \rangle_{L^2 (\Bbb T^n \times h \Bbb Z^n)}  
&=& \sum_{ k \in \Bbb Z^n}   \widehat{\psi}_1 (k)^\star \widehat{\psi}_2 (k)  + \mathcal{O}(h^\infty)  
\nonumber
\\
&=&  \langle  \psi_1 ,   \psi_2 \rangle_{L^2 (\Bbb T^n)}  + \mathcal{O}(h^\infty).  
\label{qq-dog}
 \end{eqnarray} 
In order to prove (\ref{psi-dec00}), we observe that 
\begin{eqnarray*} 
\varphi_h     = \sum_{\xi \in h \, \Bbb Z^n} \int_{\Bbb T^n}   \langle \Phi_{(x,\xi)} , \varphi_h   \rangle_{L^2 } \, \Phi_{(x,\xi)}^\star   \, dx +  \mathcal{O}(h^\infty).
\end{eqnarray*} 
is given by an $L^2$-convergent series. Thus, for any fixed $\varphi_h$ we can say that there exists $f(h) > 0$ such that
\begin{eqnarray*} 
\varphi_h     = \sum_{\xi \in h \, \Bbb Z^n, |\xi| < f(h)} \int_{\Bbb T^n}   \langle \Phi_{(x,\xi)} , \varphi_h   \rangle_{L^2 } \, \Phi_{(x,\xi)}^\star   \, dx +  \mathcal{O}(h^\infty).
\end{eqnarray*} 
$\Box$

\bigskip

\noindent
{\bf Proof of Theorem 1.2}
We apply the statement of Thm. 1, for a set of linearly independent eigenfunctions $\psi_{h,i}$ generating all the eigenspaces linked to eigenvalues $E_{h} \le E$ and $f_i (h) > 0$ given by Thm. \ref{TH1}. 
\begin{eqnarray*} 
\psi_{h,i}     = \sum_{\xi \in h \, \Bbb Z^n, |\xi| < f_i (h)} \int_{\Bbb T^n}   \langle \Phi_{(x,\xi)} , \psi_{h,i}   \rangle_{L^2 } \, \Phi_{(x,\xi)}^\star   \, dx +  R_{h,i}
\end{eqnarray*} 
where $\| R_{h,i}  \|_{L^2} = \mathcal{O}(h^\infty)$.\\
Moreover, we recall that the  Weyl Law on the number $\mathcal{N}(h)$ of eigenvalues $E_{h} \le E$ (counted with their multiplicity)  for semiclas\-sical elliptic operators (see for example \cite{G-S}) reads $\mathcal{N}(E,h) \simeq  (2\pi h)^{-n}( {\rm vol}(U(E))  + \mathcal{O}(1) )$. We define:
\begin{equation*}
g (E,h) :=  \max_{1 \le i \le \mathcal{N}(E,h)} f_i (h). 
\end{equation*}
Since any eigenfunction $\psi_h$ linked to $E_{h} \le E$ will be written as $\psi_h = \sum_{i} \langle \psi_{h,i}, \psi_h \rangle \psi_{h,i}$ then the linearity of  decomposition (\ref{deco-02}) ensures also the decomposition (\ref{deco-03}) for such $\psi_h$. Namely,
\begin{equation*}
\psi_h  = \sum_{\xi \in h \, \Bbb Z^n, \ |\xi| \le g(h,E) } \ \int_{\Bbb T^n}   \langle \Phi_{(x,\xi)} , \psi_h   \rangle_{L^2 } \, \Phi_{(x,\xi)}^\star  \, dx   +  R_h
\end{equation*}
where $R_h = \sum_{1 \le i \le \mathcal{N}(E,h)} R_{i,h}$. To conclude:
\begin{eqnarray*}
\| R_h  \|_{L^2} &\le& \sum_{1 \le i \le \mathcal{N}(E,h)} \|  R_{i,h} \|_{L^2} \le  \mathcal{N}(E,h) \max_{1 \le i \le \mathcal{N}(E,h)}   \|  R_{i,h} \|_{L^2} 
\\
&=&   \mathcal{N}(E,h) \cdot \mathcal{O}(h^\infty) = \mathcal{O}(h^\infty). 
\end{eqnarray*}
$\Box$

\bigskip

\noindent
{\bf Proof of Theorem 1.3}
We assume that $\varphi_h \in C^\infty (\Bbb T^n)$ is $L^2$ - normalized and 
\begin{equation*}
\varphi_h  = \sum_{1 \le j \le J (h)} c_j \, \psi_{h,j}
\end{equation*}
where the $L^2$-normalized eigenfunctions $\psi_{h,j}$ of $\mathrm{Op}_h (b)$ are given in Thm. \ref{TH2} and we assume $J(h) \le J_0 h^{-Q}$ for some $J_0, Q > 0$ that are independent on $0 < h \le 1$.\\ 
Define 
\begin{equation*}
\ell (h) :=  \max_{1 \le j \le J (h) } f_j (h)
\end{equation*}
where $f_i (h)$ are associated to the functions $\psi_{h,i}$ and given by Thm \ref{TH1}.\\
We now observe that if  $U_h (t) := \exp\{(- i \mathrm{Op}_h (b) t  )  / h \}$ then 
\begin{equation}
\label{UU3}
U(t) \varphi_h  =  \sum_{1 \le j \le J (h)}  c_j \ e^{- \frac{i}{h} E_{j,h} } \psi_{h,j} 
\end{equation}
for any $t \in \Bbb R$.\\
We can now apply the decomposition formula (\ref{deco-02}) with the condition on the frequencies $|\xi| \le \ell (h)$ and for  the wave function $U(t) \varphi_h$  and get the expected result, namely
\begin{equation*}
U_h (t) \varphi_h  = \sum_{\xi \in h \, \Bbb Z^n, \ |\xi| \le \ell(h) } \ \int_{\Bbb T^n}   \langle \Phi_{(x,\xi)} ,  U_h (t) \varphi_h   \rangle_{L^2 } \, \Phi_{(x,\xi)}^\star  \, dx 
 + \sum_{1 \le j \le J} R_{j,h}
\end{equation*}
for any $t \in \Bbb R$. The remainder $R_h := \sum_{1 \le j \le J} R_{j,h}$ can be estimated as in the previous Theorem, namely
\begin{eqnarray*}
\| R_h  \|_{L^2} &\le& \sum_{1 \le i \le J} \|  R_{j,h} \|_{L^2} \le J_0 \, h^{-Q}  \max_{1 \le j \le J}   \|  R_{j,h} \|_{L^2} 
\\
&=&  J_0 \, h^{-Q}  \cdot \mathcal{O}(h^\infty) = \mathcal{O}(h^\infty). 
\end{eqnarray*}
$\Box$

\end{document}